\documentclass[preprint,aps]{revtex4}

\usepackage{graphicx}
\usepackage{dcolumn}
\usepackage{bm}
\def\ä{\"{a}}
\def\ü{\"{u}}
\def\ö{\"{o}}
\def\Ä{\"{A}}
\def\Ü{\"{U}}
\def\Ö{\"{O}}

\begin{document}

\preprint{APS/123-QED}

\title{Device for \emph{in-situ} cleaving of hard crystals}

\author{M. Schmid}
\email{martina.schmid@physik.uni-augsburg.de}
\author{A. Renner}
\author{F. J. Giessibl}


\affiliation{%
Universit\"at Augsburg, Institute of Physics, Electronic
Correlations and Magnetism, Experimentalphysik VI,
Universit\"atsstrasse 1, D-86135 Augsburg, Germany.
}%

\date{Submitted to Review of Scientific Instruments, originally received May 31 2005, revision 2}

\begin{abstract}
Cleaving crystals in a vacuum chamber is a simple method for
obtaining atomically flat and clean surfaces for materials that
have a preferential cleaving plane. Most \emph{in-situ} cleavers
use parallel cutting edges that are applied from two sides on the
sample. We found in ambient experiments that diagonal cutting
pliers, where the cleavage force is introduced in a single point
instead of a line work very well also for hard materials. Here, we
incorporate the diagonal cutting plier principle in a design
compatible with ultra-high vacuum requirements. We show optical
microscopy (mm scale) and atomic force microscopy (atomic scale)
images of NiO(001) surfaces cleaved with this device.
\end{abstract}

\maketitle
\section{Introduction}
Surface science relies on the availability of atomically clean and
well-defined surfaces. Cleaving a crystal in an ultra-high vacuum
environment is a simple means to provide clean and atomically flat
surfaces. Existing implementations of cleavers use a spring-loaded
blade that moves towards an anvil
\cite{Pela:1973,Janssen:1974,Angot:1991} with parallel edges, an anvil
with one blade that presses against the sample that is to be cleaved
\cite{Dupoisson:1976,Carr:1988,Yates:1998} or a device that pushes the
sample against an obstruction
\cite{Giessibl:1991b,Ohta:1994,Kim:1996}. Other techniques use one
blade that is introduced from one side of the vacuum chamber with a
wobble-stick type feedthrough hitting a sample that is countered by a
fixed or movable anvil \cite{Reichling}. Cleaving crystals \emph{in
situ} is relatively simple for soft materials, such as alkali halides.
However, cleaving hard materials as NiO is more difficult. We could
successfully cleave a strip cut from a NiO wafer with a cross section
of roughly 2\,mm by 0.6\,mm with the techniques described in
references \cite{Giessibl:1991b,Ohta:1994,Kim:1996}. However, in
particular on hard samples, the sample areas close to the crystal
boundaries show very poor surface quality and it is desirable to
provide cleavage faces with greater lateral dimensions to obtain good
quality surface areas. We found in ambient experiments that cleaving
NiO with a razor blade that touches the sample along a line is rather
difficult, while cleaving it with a wire cutter is simple and requires
little force even for large cross sections. In a wire cutter, the
cleavage force is introduced in a single spot. This highly localized
stress field is known to facilitate cleaving \cite{Eberhart:2003}. We
therefore aimed to incorporate this principle in a vacuum compatible
form, shown in the next section.

\section{Experimental implementation}

Our cleaver (see Fig. 1 (a)) sits on a single CF\,35 flange that
holds a rotary feedthrough capable of transmitting a torque of
10\,Nm mounted on a CF\,16 flange. The rotary feedthrough connects
to an axle that has the cleavage knife mounted to it at its end.
The CF\,35 flange has a pipe with venting holes welded to it,
holding an anvil at its end and housing a ball bearing that
supports the axle. Because the knife edge is not parallel to the
surface when the crystal is not cleaved yet, the cleavage force is
applied at a well-defined point (see Fig. 1 (b)), facilitating the
cleavage process and usually resulting in fairly flat cleavage
planes. When the knife edge is touching the crystal along its
whole length, as done in many previous cleavage designs, the
cleavage can start at multiple points that are on different
parallel crystal planes (see Fig. 1 (c)). The rotating blade and
the steady anvil resemble the two cutting blades of a wire cutter.
In the present cleaver, the blade is made of stainless steel, but
it could also be manufactured from hardened steel or other hard
materials such as tungsten carbide. The blade angles are chosen
such that once the cleave is initiated, the blade pushes the
sample backwards and the blade does not scratch over the freshly
cleaved surface (Fig. 1(c)). The maximum force that can be applied
to the initial point of contact between knife and sample is given
by the length of the arm and the maximal torque rating of the
rotary feedthrough. In our setup, it is approximately 1\,kN.

\section{Results}
Figure 2 shows an optical microscopy image of a typical cleavage
plane. The steplines clearly merge into the point where the
initial cleaving force was applied. The cleavage process is shown
in real time for soft (KBr) and hard (NiO) materials by two
movies, online available at www.xxx.yyy. The advantage of a large
cleavage area is that high-quality surface areas are more likely
to be found. In most experiments, it is not possible to obtain
atomic resolution over the whole surface area (see caption Fig.
2). Debris on surfaces is especially harmful for scanning probe
microscopy studies, because the radii of the probe tips are often
large and even small pieces of debris on otherwise perfectly flat
and clean surfaces can prevent the tip apex from reaching the flat
surface, because other tip sections may make contact with the
debris before the apex reaches the sample.

With the cleaver shown in Fig. 1, we could easily cleave cross
sections on the order of 2 by 4\,mm$^2$ and we could find surfaces
that are clean enough to allow atomic resolution by AFM in
approximately 9 out of 10 cleavage trials. Figure 3 shows typical
atomic force microscopy images of an \emph{in situ} cleaved
NiO(001) surface that was produced with this device.

Because the cleaver can be mounted on a single flange, it is very
compact. As it uses only a single rotary feedthrough and few other
parts, it is easy to build and quite cost effective.

\section{Acknowledgments} We thank Jochen Mannhart for support and useful suggestions,
Marilyn Gleyzes for assistance with preparing the mechanical
drawings, Rainer P\ätzold for machining the second version of the
cleaver and Michael Reichling for helpful comments. This work is
supported by the Bundesministerium f\"{u}r Bildung und Forschung
(project EKM13N6918).

\newpage
\begin{figure}
  \includegraphics[width=14cm]{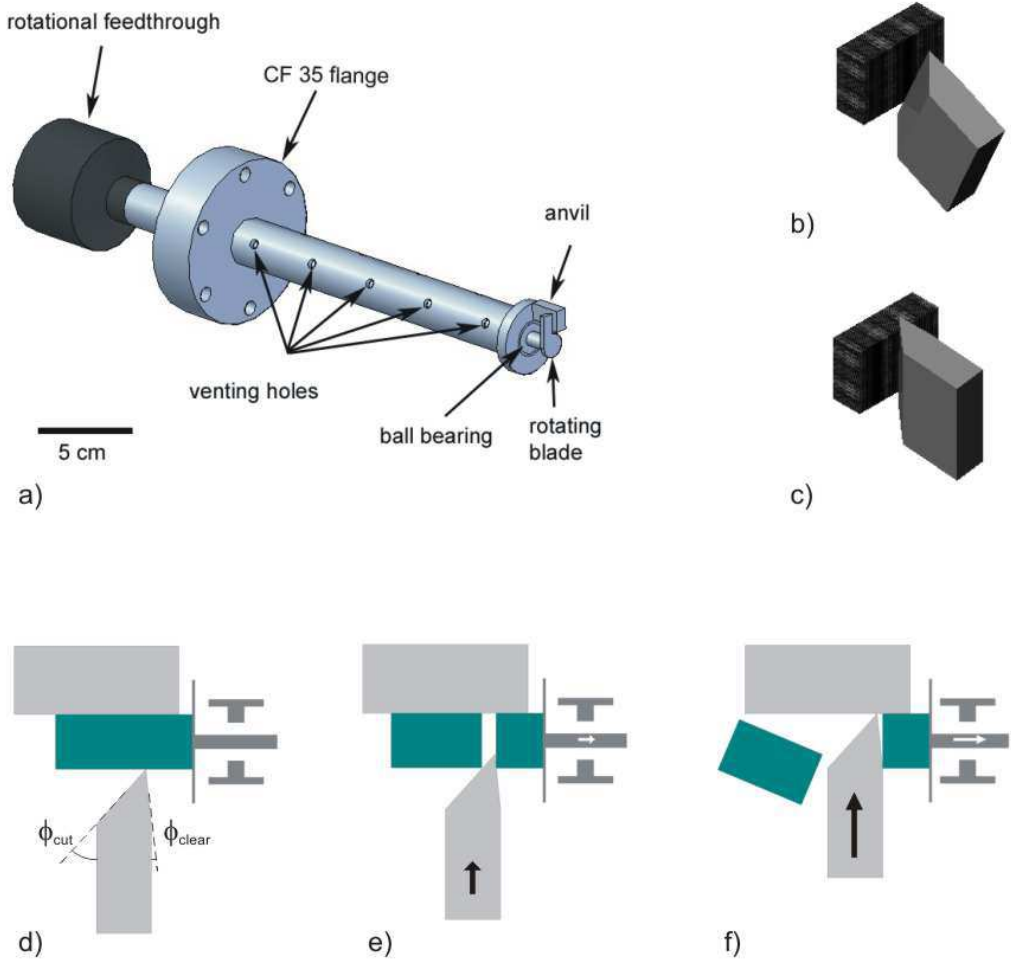}\\
  \caption{(a) Perspective view of the single-flange cleaver. It
  consists essentially of three parts: a rotary feedthrough,
  an axle holding the cleavage knife that is connected to the rotary feedthrough and a tube
  with venting holes that has the anvil welded to it. (b) Cleaving
  a crystal using the wire cutter principle results in a well-defined point where the cleave is initiated. In contrast to a cleaver
  where the blade is applied parallel to the side faces of the crystal sample to be cleaved (c),
  the wire cutter principle does not require perfect alignment
  between the cutting edge and the intended cleavage direction. (d)
  Schematic view of the blade with cutting angle $\phi_{cut}$ and
  clearance angle $\phi_{clear}$ in the initial phase of cleavage. In our setup, $\phi_{cut}\approx
  42^{\circ}$ and $\phi_{clear}\approx -6^{\circ}$. (e,f) The negative
  clearance angle ensures that the cleavage plane remains
  unscathed when the blade slides over the sample surface after
  the cleave (see also online movies). The sample holder is movable along a line
  perpendicular to the cleavage plane, enabling the sample to move backwards (white
  arrow in (e) and (f)).}
\end{figure}

\begin{figure}
  \includegraphics[width=12cm]{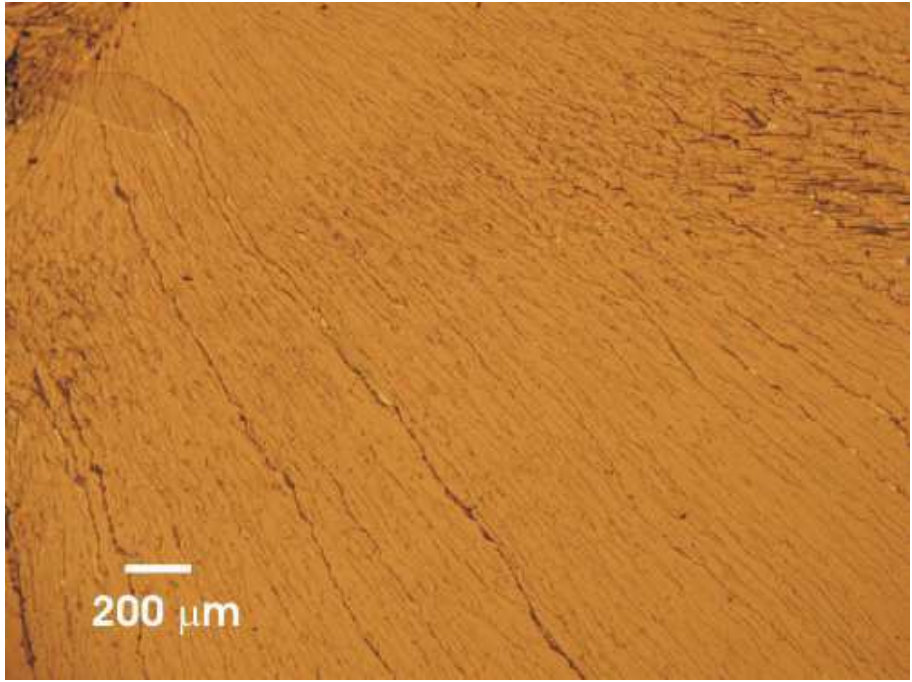}\\
  \caption{Optical microscopic view of the cleaved NiO(001)
  surface. Cleavage was initiated on the top left corner of the
  sample. The central region of the cleavage plane has a rather
  good surface quality, while the sections close to the boundaries
  are more rugged and are sometimes covered with debris.}
\end{figure}

\begin{figure}
  \includegraphics[width=8cm]{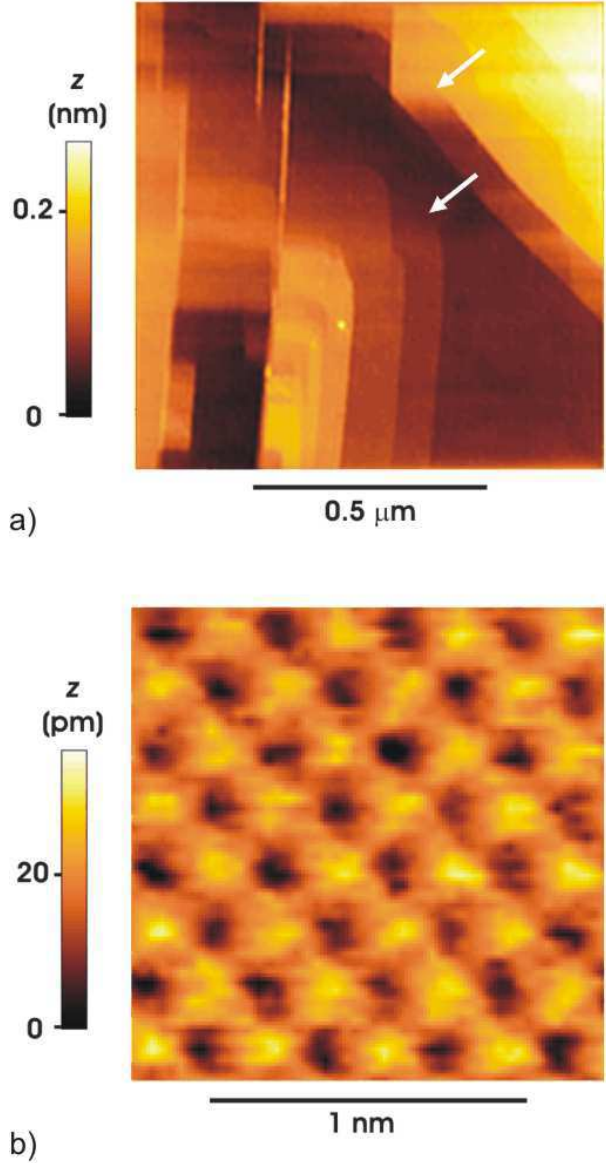}\\
  \caption{Atomic force microscopy images of the cleaved NiO(001)
  surface. NiO(001) has a rocksalt structure with a cubic lattice constant
  of $a_0=417$\,pm and natural step heights of integer multiples of $a_0/2$.
  (a) Image size 1\,$\mu$m$\times$1\,$\mu$m, showing atomic steps with heights of
  $a_0/2$ and $a_0$. At the top, the NiO crystal is 3 unit cells higher at the right edge than
  at the left, and at the bottom, the left edge is 3 unit cells higher than the right edge.
  The misorientation angle of the actual cleavage plane with respect to the ideal (001) plane
  is thus on the order of 3\,$a_0/1$\,$\mu$m\,rad = 0.07$^\circ$. The image shows that the (001)
  surfaces are not ideal -- a few screw dislocations are present (arrows).
  (b) Atomic resolution image (size 1.5\,nm$\times$1.5\,nm). Images of this quality could only be obtained on
  surface areas that are flat and free of debris for fairly large
  areas. The larger the area of a cleaved surface, the more likely
  it is possible to find such good surface areas.}
\end{figure}

\end{document}